\documentclass[lnicst]{svmultln}
\usepackage[pdftex]{graphicx}
\usepackage{amssymb,amsmath}
\usepackage{subfigure,cite}

\usepackage{color}

\usepackage{makeidx}  
\begin{document}

\mainmatter              

\title{Micro Service Cloud Computing Pattern \\ for Next Generation Networks}

\author{Pascal Potvin\inst{1,2}, Mahdy Nabaee\inst{1,3}, \\ Fabrice Labeau\inst{3},
Kim-Khoa Nguyen\inst{2}, Mohamed Cheriet\inst{2}
 }

\authorrunning{Potvin \textit{et al.}}   

\institute{Ericsson Canada Inc., Montreal, Canada\\
\and Ecole de Techologie Superieure, Montréal, Canada \\
\and ECE Dept, McGill University, Montreal, Canada \\
\email{{pascal.potvin, mahdy.nabaee}@ericsson.com} \\
\email{fabrice.labeau@mcgill.ca}, \email{knguyen@synchromedia.ca}\\\email{mohamed.cheriet@etsmtl.ca} 
}

\maketitle             

\begin{abstract}

The falling trend in the revenue of traditional telephony services has attracted attention to new IP based services.
The IP Multimedia System (IMS) is a key architecture which provides the necessary platform for delivery of new multimedia services.
However, current implementations of IMS do not offer automatic scalability or elastisity for the growing number of customers.
Although the cloud computing paradigm has shown many promising characteristics for web applications, it is still failing to meet the requirements for telecommunication applications.
In this paper, we present some related cloud computing patterns and discuss their adaptations for implementation of IMS or other telecommunication systems.


\end{abstract}

\section{Introduction and Motivation}
\label{sec:intro}

{The increasing demand for telecommunication services has made the providers to invest further in their infrastructure.
The cost of upgrading the infrastructure as well as the competition between different providers is resulting in falling revenue obtained from traditional telephony services. 
This fact has led the providers to look for other revenue sources by offering new multimedia services.
However, the rising number of clients and their data usage is increasing the traffic load on the core of telecommunication networks which requires high cost provisioning of the network.}


\begin{figure}[t]
\centering
\resizebox{.51\textwidth}{!}{
\includegraphics{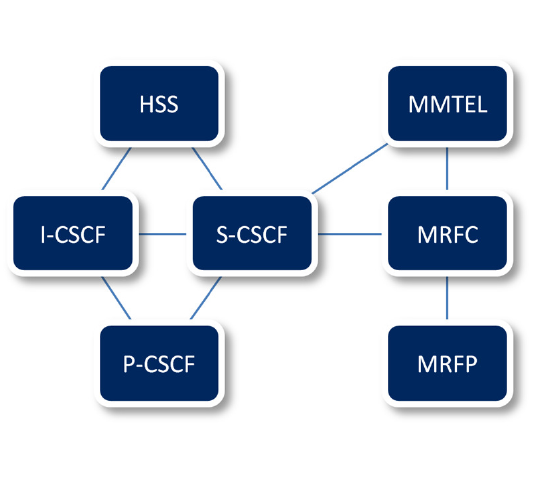}}
\caption{\label{fig:ims}The main functionalities of the IP Multimedia System: Call Session Control Function (CSCF), Home Subscriber Server (HSS), Multimedia Telephony (MMTEL), and Media Resource Function (MRF).
}
\end{figure}

As the main path toward the next generation network, IP multimedia subsystem (IMS) is an architectural framework for end-to-end delivery of multimedia services via IP-based mechanisms \cite{poikselka2013ims}. 
The IMS is built upon Session Initiation Protocol \cite{handley1999sip} and Real-time Transfer Protocol \cite{schulzrinne1996rtp} for control and data planes, respectively.
As it is shown in Fig.~\ref{fig:ims}, the main modules in an IMS include Call Session Control Functions (CSCF), Home Subscriber Server (HSS), Multimedia Telephony (MMTEL) and Media Resource Functions (MRF).

The CSCF is the core function in an IMS system, which is in charge of performing the appropriate signaling between the user equipment (UE) and IMS modules.
Further, the CSCF handles the establishment and termination of sessions, authentication, security and Quality of Service monitoring.
Depending on the specific task of a CSCF unit, it is divided in to Proxy (P), Interrogating (I) and Serving (S) types of CSCF, as shown in Fig.~\ref{fig:ims}.

HSS is the main database unit in IMS, which keeps the profile of all subscribers and the necessary triggers for their policies.
MMTEL unit enables end-to-end real time services between the parties for different multimedia contents including real time video, text messaging and file sharing.
Finally, MRF (usually divided in control and processing modules) is in charge of delivery of the media services by providing media related functions such as voice mixing for voice content.

Although different resolutions of IMS have been developed for commercial use in the industry, its efficiency and low cost delivery needs further investigation. 
Especially, the main drawbacks of current IMS infrastructure are manual (human based) scalability\footnote{They rely on human operations to deploy further resources to accommodate the increase of demand.}, lack of elastisity and high deployment and maintenance costs.

Thanks to virtualization techniques, sharing of computing, storage and network resources has been made possible, resulting in the creation and growth of cloud computing \cite{buyya2013mastering}. 
By abstracting the hardware and software, Infrastructure as a Service (IaaS) provides a pool of computing and storage resources which isolates us from the complexity of dealing with individual hardware devices. 
Meanwhile, since many cloud users have access to these shared computing resources, they can change their subscription volume, resulting in an elastic behavior.
Current architectures of cloud computing are designed to provide the best services possible and are failing to provide any telecommunication-level quality of service (QoS) assurances \cite{glitho2014cloudifying}.




Research on the implementation of IMS core network within the cloud computing infrastructure is in its early stage and there are few competitive published work. 
The 3GPP standardizing body has attempted to design the IMS such that its main functionalities (especially Call Session Control Function) are to some extent scalable \cite{glitho2014cloudifying} although this limited scalability does not translate in elasticity.
A few other works have focused on the scalability of the individual functional units of IMS; for instance, the authors in \cite{yang2011new} have addressed the scalability of Home Subscriber Server (HSS) with the aid of concepts in distributed databases.
In \cite{lu2013virtualization}, the authors have proposed a resource allocation which satisfies the time requirements at the level of a telecommunication network. 
They have accomplished this by using static and dynamic groups for assignment of virtual computation units to the physical computation units. 

In this paper, we introduce a new architecture for elastic implementation of IMS which is based on micro services.
In contrast with previous work, our new architecture can be implemented on top of different cloud or node based computing services including IaaS and PaaS (Platform as a Service).
Further, we propose a mechanism to trigger the allocation of new computing nodes to accommodate overloaded nodes. 
This will enable us to have automatic scalability and achieve elasticity for the implementation.

In section~\ref{sec:computingSchemes}, we describe some of the cloud computing patterns and their application for the IMS. Specifically, we discuss micro service architecture and describe our micro service based architecture for IMS.
In section~\ref{sec:loadBalancing}, we describe the load balancing mechanism used in our architecture which is followed by our discussion on the automatic scalability of our architecture in section~\ref{sec:autoScale}.
We present some of our experimental results using our proof of concept implementation in section~\ref{sec:results}.
Finally, we present our concluding remarks and future works in section~\ref{sec:conclusion}.





\section{Cloud Computing Scaling Schemes for IMS}
\label{sec:computingSchemes}

One of the main challenges of using cloud architectural patterns is to adapt stateless web technologies for the strictly stateful telecommunication applications.
Specifically, we need to adopt mechanisms which enables us to use the current cloud architectures for telecommunication applications with a lot of state information, \textit{e.g.} the state of SIP handling in IMS.
Moreover, we need to study the relation between the cloud related metrics (\textit{e.g.} load of the units) and the telecommunication related metrics (\textit{e.g.} Quality of Service).
In the remaining of this section, we study some of the conventional architectures for the cloud and discuss their adaptation for our IMS implementation.



In the literature, scaling an application is categorized as three different axes. 
Running multiple instances of the whole application using a load balancer is referred to as the \textit{x} axis of scaling.
The \textit{y} axis stands for the splitting of the application into smaller components where each component is a service responsible for a specific functionality.
Finally, in the \textit{z} axis of scaling, the input data are partitioned into different segments where the segments are handled by different computing resource (also called \textit{sharding}).
%

\subsection{Micro Service Architecture for IMS}
\label{sec:microService}

In the \textit{y} axis of scaling, the application is decomposed into smaller units, called \textit{micro services}.
Decomposing into smaller micro services will enable us to distribute the computational load of the application among different hardware devices or even geographical locations.
This will provide the management of computational load and resources with a fine granularity.
Further, the micro services architecture will provide a flexibility for deployment by allowing the micro services to be deployed at different computational resources, \textit{e.g.} different virtual machines, platforms, containers or even geographical locations.

On the other hand, having an architecture which is built up from smaller services makes parallel and continuous software development feasible.
The ability to reuse micro services for other applications is another advantage of this pattern.
Specifically, a micro service implements a small set of functionalities which can be interfaced with other modules of the application via synchronous or asynchronous communication protocols (\textit{e.g.} TCP and UDP).
Such an architecture will enable us to deploy (execute) a micro service anywhere without being bound to a specific computing node and therefore achieve a fully scalable architecture.



\begin{figure}[t]
\centering
\resizebox{.94\textwidth}{!}{
\includegraphics{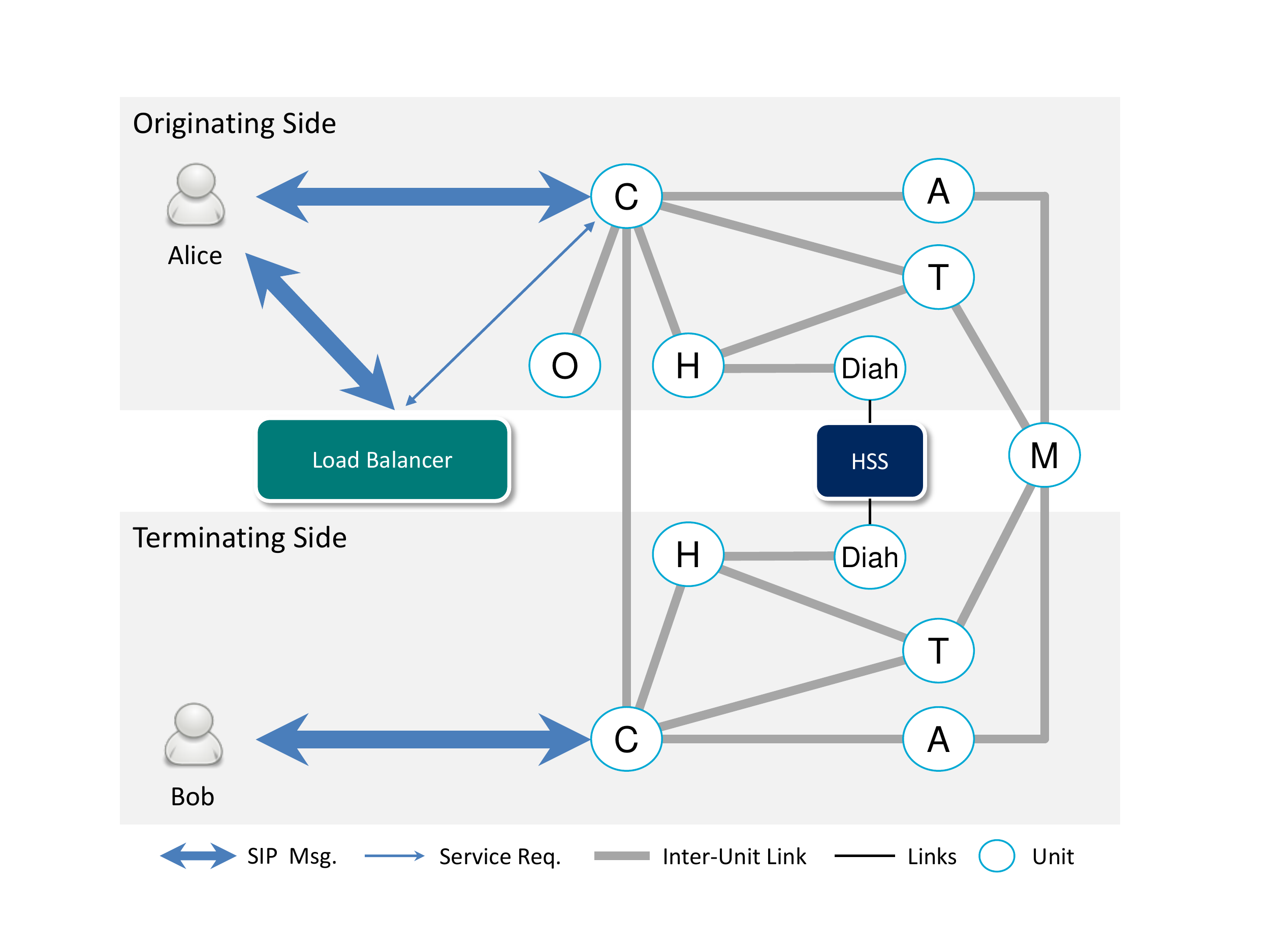}}
\caption{\label{fig:ims_2}Micro service implementation of IP Multimedia System: The main functions of IMS are split into smaller micro services, called CSCF (C), Orchestrator (O), Anchor point (A), Telephony server (T), Media processor (M) and HSS front end (H). 
This allows us to dynamically create micro services for each IMS subscriber or call session.
}
\end{figure}

As shown in Fig.\ref{fig:ims_2}, we have divided the IMS functions into smaller micro services where each micro service runs on a \textit{dynamically allocated} computing resource. 
To establish a call, the originating side will send its INVITE to the load  balancer (as the entry point of our IMS).
The load balancer will then determine a computing node and create an instance of the C actor (CSCF) to interact with the originating side and handle the call session functions.
The Orchestrator (O) is then created by the C and will determine the subscription details for that call session, depending on the subscription policy.

As it is shown in Fig.\ref{fig:ims_2}, the H unit is in charge of interacting with the HSS to fetch and update the user profiles via the Diameter stack protocol.
The A (Anchor point controller) and T (Telephony server) are in charge of creating and updating the call session settings.
Finally, the M unit is the media processor which processes the content of the call sessions (\textit{e.g.} bit rate matching for voice calls).\footnote{Although the described micro service architecture does not include all of the functionalities in an IMS, it povides a good coverage for most of its functionalities and makes us able to study different aspects of cloud based implementation of IMS, as discussed in this paper.}

Splitting the IMS architecture into small units (called micro services) provides us with a lot of flexibility on the deployment of the units.
Explicitly, each of the units involved in the establishment of a call session (shown in Fig.~\ref{fig:ims_2}) can be deployed on a different computing node as they interact together via TCP communication protocol.
Hence, the fine granularity offered by our architecture makes it easier to scale the system with the demand (\textit{i.e.} increasing and decreasing the amount of allocated computation resources with the new and terminated call sessions).
However, it should be noted that such advantage is achieved by having higher latency in the IMS functionalities.\footnote{In section~\ref{sec:results}, we discuss this further by presenting our experimental results for the latency.}


\subsection{Computing Nodes as Pouches}
\label{sec:pouches}

The virtualization platforms have provided us with an abstract encapsulation of  computing, storage and network resources.
From the perspective of applications (\textit{e.g.} IMS micro services), this encapsulation is similar to a physical computing node like a blade server.
However, in practice, it may be a \textit{virtual machine} in OpenStack platform or a \textit{container} in Continuum platform \cite{apcera}.
In general, we call a set of physical or virtual computing resources which are isolated from other sets of resources (for the application), a \textit{pouch}.

The key point in our IMS implementation is the separation of the platform in which a pouch is deployed (or instantiated) and the application micro services.
In other words, each of the micro services (discussed in the preceding section) can be run (executed) on a different (type of) pouch, independent of the pouch platform.

%
%
%

\subsection{Horizontal Scaling and Sharding Pattern}
\label{sec:sharding}

As the main requirement for scalability, the applications are supposed to work such that adding more computing resources would be the solution to handle further queries, calls or subscriber.
A load balancer will then be able to spread the computational load on these computing resources in a way to achieve the quality requirements.
Such an architectural pattern has reached its maturity for current web services.
However, the stateful nature of telecommunication applications requires further adaptation of native web technologies and services for use in telecommunication applications.


\begin{figure}[t]
\centering
\resizebox{.96\textwidth}{!}{
\includegraphics{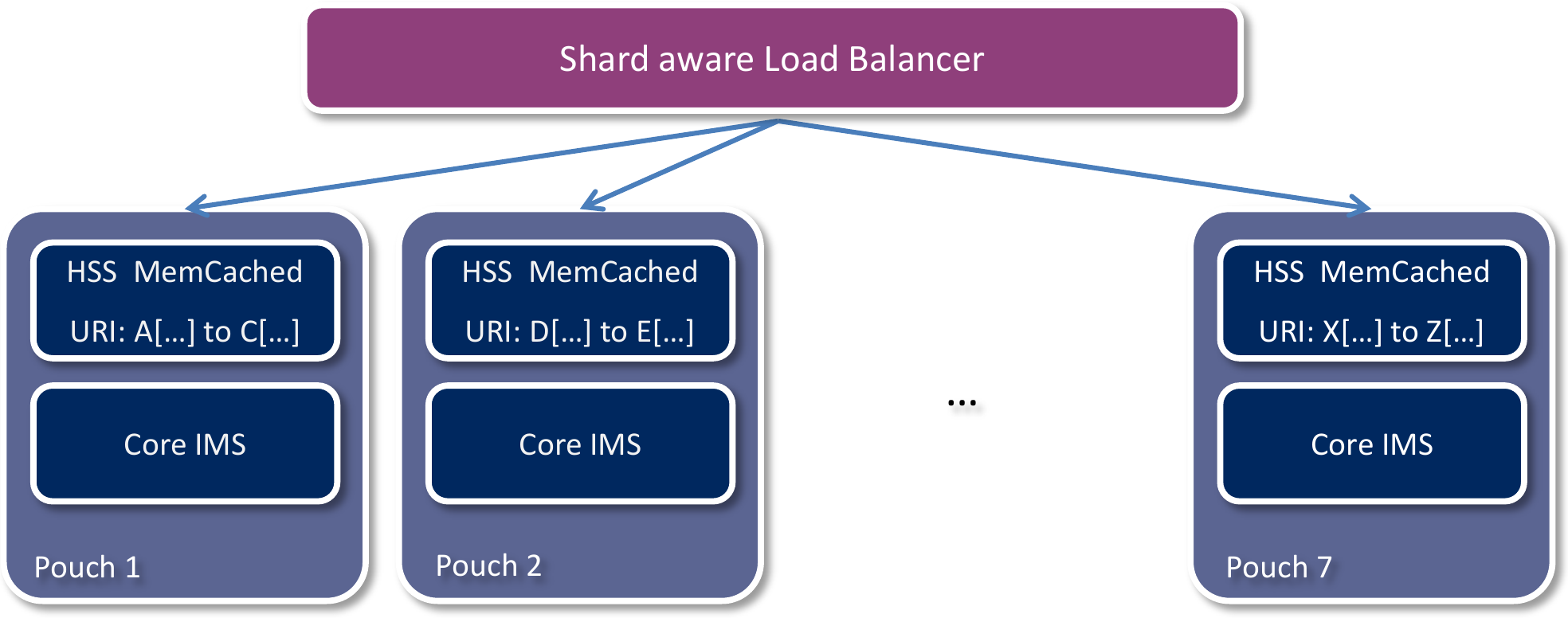}}
\caption{ Sharding of Database: The content of HSS is divided into multiple local caches at each pouch to decrease the query load on the main HSS.
The subscribers are (as much as possible) directed to the same pouch which their profile is cached.
\label{fig:sharding}
}
\end{figure}

One of the key approaches in our design of IMS is based on the concept of sharding where the user database is split into a number of shards (databases).
Essentially, at each shard of our IMS architecture, a partition of the user data with the same key is stored (cached).
Further, each of these shards are also kept consistent (synchronized) with a centralized HSS entity which results in a smaller number of queries (smaller load) on HSS.
Moreover, the key used in the sharding is a hash function of the call session identifier and may be as simple as hard division of first letter of the SIP (Session Initiation Protocol) identifiers.

The proper design of sharding can help us achieve horizontal scaling and makes us able to recover from computation node failures by creating the cache on a new node.
Further, the load balancer will be able to distribute data such that nodes are loaded almost uniformly.
The load balancing algorithm used for our IMS architecture is explained in the following.

\section{Load Balancing and Scaling}
\label{sec:loadBalancing}

The load balancer goal is to distribute the computational load between the computing nodes and decrease the caching load from HSS.
In the following, we describe our proposed load balancing mechanism for IMS.

\begin{figure}[t]
\centering
\resizebox{.99\textwidth}{!}{
\includegraphics{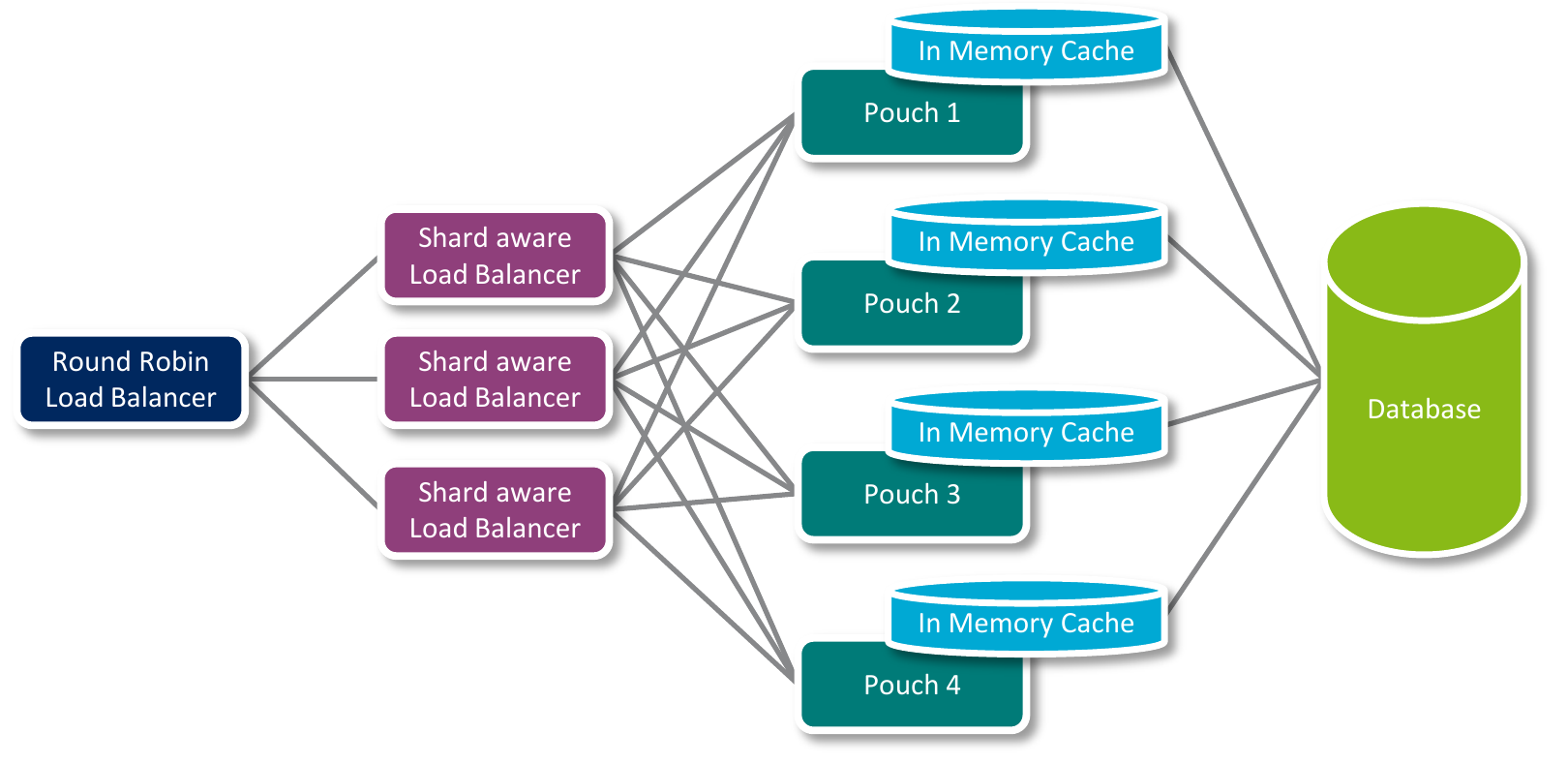}}
\caption{ Rendez-vous Load Balancer and Database Sharding: The content of HSS is divided into multiple local caches at each computing node to decrease the query load on HSS.
The subscribers are (as much as possible) directed to the computing node where the cache has their profile.
\label{fig:rendezvousLB}
}
\end{figure}

When a new SIP message is received at the load balancer (entry point in our IMS), it is sent to a randomly picked \textit{rendezvous} load balancer (for example using a round robin mechanism).
As shown in Fig.~\ref{fig:rendezvousLB}, the rendezvous load balancer is then in charge of assigning the computing nodes for handling of SIP message.

The rendezvous load balancers find the computing node by using the highest random weight mapping, introduced in \cite{rendezvousLB}.
Specifically, for each computing node, the hash of the  combination of originating side URI and computing node host name (or IP address) is calculated.
The combination can simply be the concatenation of the originating URI and host name strings.
The computing node with the highest (or alternatively lowest) hash value is then selected for caching of subscriber profile and handling of the call:
\begin{equation}\label{eq:highest}
\mbox{selected~node} = \arg\max_{\mbox{node}~i} \mathcal{H} ( <\rm{URI}> + <\rm{host}_{i}>)
\end{equation}
In \eqref{eq:highest}, $\mathcal{H}(\centerdot)$, $<\rm{URI}>$ and $<\rm{host}_{i}>$  represent the hash function, originating side URI and host name of the computing node, respectively.
Moreover, $+$ denotes the concatenation operation for two strings.

Since our load balancing algorithm only uses the subscriber URI, all of the calls from the same subscriber are directed to the same computation node.
As a result, its user profile does not need to be fetched from the main HSS entity again and can be read from the local cache of that computation node.

Conventionally, multiple instances of HSS with different user profiles were deployed in IMS and 
a Subscriber Location Function (SLF) was used to direct the calls to their specific HSS unit \cite{poikselka2013ims}. 
In contrast with the SLF mechanism, our approach is designed to minimize the number of database queries out of a pouch.
Although the SLF decreases the load on the HSS units, one may still need to make database queries out of the pouch where the computation is done.
Further, since the local database queries can be made faster than the external ones (as it does not need to be done via network), our local caching mechanism should be a better alternative for long term handling of a set of stationary subscribers.

%
%

In the mentioned algorithm, the rendezvous load balancer may find a computing node which is overloaded and can not handle a new subscriber.
In such cases, a busy signal is sent to the management unit and the call session is dropped.
Essentially, the management unit is in charge of preventing such cases by provisioning and monitoring of the computing resources and allocating new computing nodes, as discussed in the next section.


%
\section{Automatic Scaling and Busy Signal}
\label{sec:autoScale}

As it was described above, the load balancer will notify the management unit to create new computing nodes if a node is overloaded to handle a new subscriber or call.
However, the metrics used to determine the load of a computing node in web services is different from those in telecommunication applications.
The primary and final goal in telecommunication applications is to satisfy the Quality of Service (QoS) requirement of the service; \textit{e.g.} latency for establishing calls.
Hence, one needs to find an appropriate mechanism to track and predict the changes in QoS for the subscribers.

In our design for IMS, we implemented a \textit{busy signal} mechanism for this purpose which results in dynamic allocation of pouches.
Essentially, when a subscriber is directed to a computing node by the load balancer and that node is overloaded, a busy signal would be sent to the management unit and the call session is dropped.
Then, the management unit (which can be part of the load balancer functionality) will create a new computing unit and update the record of computing nodes at the rendezvous load balancers.
The uniform nature of the hash function ensures that the loads of the computing nodes are distributed almost uniformly.
Further, this mechanism prevents the subscribers from being directed to different locations which would result in a large number of queries to the HSS.

The measurements used to determine if a computing node is overloaded have to be carefully chosen and this is still a subject of study in the literature.
Notably, the conventional metrics used for measuring the computational load of a node (\textit{e.g.} processing unit load and memory usage) are not directly applicable to the QoS in IMS.
Furthermore, the QoS between two subscribers depend on the communication link and network where the traffic travels through.
In cases where a computing node is located in a congested part of the network where delivery time of packets is not acceptable, the load balancer will create a new computing node and direct the new subscribers and calls to it.

Although there is still no specific formula to draw QoS conclusions from different metrics representing the condition of the computing nodes, our empirical experiments showed that one has to consider the following to determine if new pouches need to be created:
\begin{itemize}
\item the round trip transmission delay between geographically distributed computing nodes and user equipment,
\item the processing load (\textit{e.g.} processing unit and memory usage) of individual computing nodes,
\item the history of the QoS experienced by different subscribers and their subscription policies.
\end{itemize}

\section{Experimental Results}
\label{sec:results}

In this section, we present some of our experimental results, obtained from our implementation.
Most of the units in the IMS architecture (shown in Fig.~\ref{fig:ims_2}) are involved in the establishment of the call sessions.
After the call is established, the content of the call (\textit{e.g.} voice) is handled by the media processor (M) unit.
Therefore, it need to have a very small number of interactions with the other units which may affect the computational performance of the media processor.
As a result, we focus on the call establishment process and evaluate the performance of our architecture in terms of the latency in the establishment of a call.


We have implemented and deployed our micro service IMS application on a cluster of Raspberry Pi, as a proof of concept. Our cluster is built of eight model B Raspberry Pi \cite{pi} which are put together similar to a cabinet of blade servers. 
Each Raspberry Pi is considered as a pouch and they are linked together by using an Ethernet switch.
One of the pouches is used to host the load balancer as the entry point of the SIP messages which is used by the SIP client at the beginning of their SIP signaling.

\begin{figure}[t]
\centering
\resizebox{.99\textwidth}{!}{
\includegraphics{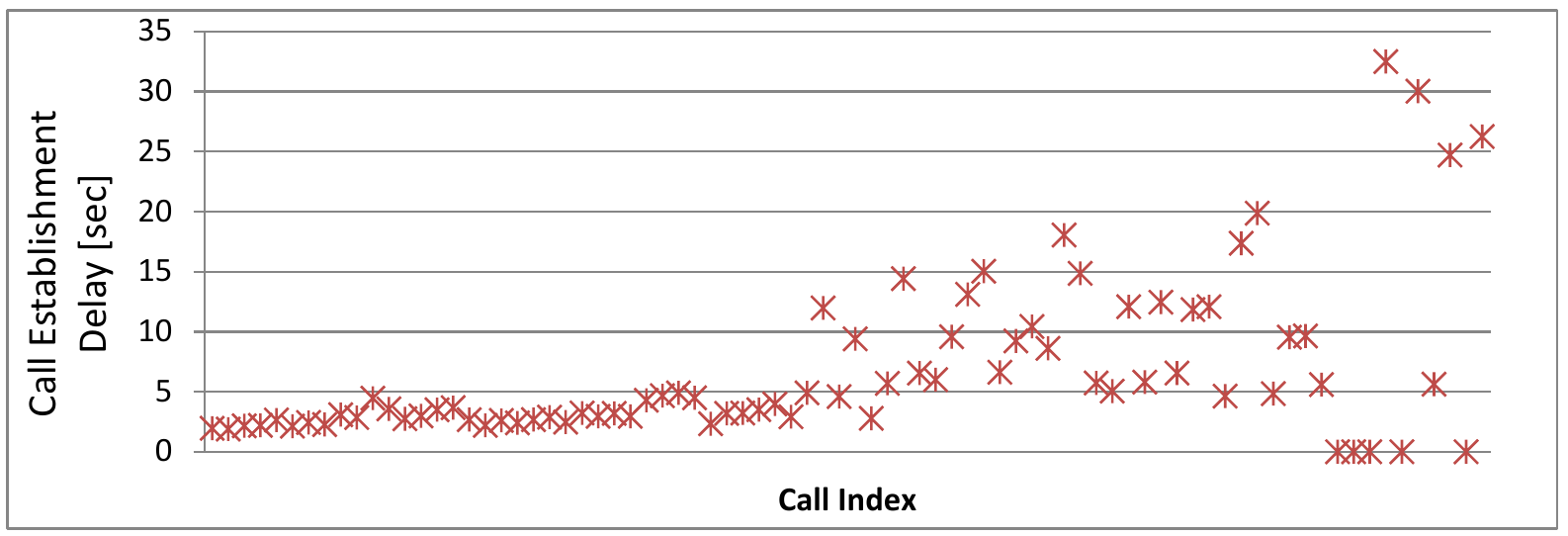}}
\caption{Call establishment latency for different calls: The latency increases when the load on the system (number of concurrent calls) increases.
\label{fig:result}
}
\end{figure}

Our experiments are done by making a number of SIP calls via our IMS implementation.
Specifically, we put a maximum of $80$ concurrent calls to the system with a constant rate of $30$ calls per minute. 
After a call is established, it lasts $300$ seconds and then it is terminated. 
The calls are made between different pairs of subscribers and only carry voice content. 
In Fig.~\ref{fig:result}, we have depicted the call establishment latency for each call.
In this figure, the vertical axis represents the delay in seconds and the numbers on the horizontal axis correspond to the order in which the calls are placed.

As it is shown in Fig.~\ref{fig:result}, the call establishment latency is around $2$ seconds which is in a comparable range with the the results in \cite{boumezzough2013evaluation}.
However, as the number of concurrent calls on the system increases, the load on the system is increased which results in higher call establishment delays.
Especially, this happens for calls which are initiated late during our experiments.
The high latency values for the establishment of late calls is mainly because the pouches are overloaded by the actors corresponding to the other previous calls.
Specifically, the concurrent C, A and T processors, created for different call sessions, put a lot of load on Raspberry Pi's which may result them not to be able to handle the new call establishments with a reasonable latency.

\section{Conclusion and Discussion}
\label{sec:conclusion}

In this paper, we discussed some of the cloud computing patterns and studied the possibility of applying them to the IP multimedia system.
Specifically, we described a new cloud based architecture for implementation of IMS where the system is split in a number of micro services.
Such micro service based architecture made us able to adopt automatic scalability to achieve elasticity for the IMS.
%
We also proposed to use local caching as an effective approach to reduce the number of queries to the main HSS unit which is a major bottleneck in the IMS architecture.
Further, we discussed rendezvous load balancing in order to achieve uniform load distribution among the computing nodes and reduce the communication overhead by handling the calls of a subscriber at the same computing node (and take advantage of local caches).
In the following, we discuss some of the important aspects of our work in progress.

\subsection{Computing Node Failure}
\label{sec:nodeFailure}

Traditional implementation of telecommunication functionalities on dedicated hardware does not provide a mechanism for automatic recovery and migration.
As a result, the recovery of failed devices only has to rely on manual (human assisted) maintenance and operation.
This fact requires the telecommunication equipment to have a very high \textit{mean time between failures}.

The isolation of pouches and the IMS micro services allows us to move the deployed micro services to a new instantiated pouch in case of a failure.
This advantage of cloud computing has introduced a change in the definition of reliability for the systems.
Since the cloud platform provides us with a large pool of pouches (or virtualized computing nodes), the frequency of happening of a failure is not a critical factor for cloud based implementations.
However, the new mindset requires a low \textit{mean time to recovery} for the new cloud based implementations of telecommunication applications.
Specifically, it is now important to be able to create new pouches and move the functionalities of the failed pouch (or device) to the new create pouches.

It is fair to say that this new mindset is mainly due to the advance of \textit{stateless} web services on the cloud.
However, implementing \textit{stateful} telecommunication applications on top of fully stateless pouches has many technical challenges and may result in higher latency.


\subsection{Generic Management Unit}
\label{sec:mano}

In section~\ref{sec:autoScale}, we discussed automatic scaling using application related parameters (\textit{e.g.} QoS) to achieve elastisity in our architecture.
Isolation of the IMS application functionalities from the cloud management unit is one of the interesting features in development of generic cloud management with automatic scalability for telecommunication applications.
To develop such generic architecture, we need to find an appropriate figure which is independent of the application layer and is capable to reflect the level of used resources by the application.
Having such generic figure, the management unit will be able to increase or decrease the number of pouches, allocated for the application with the changes of the load (\textit{e.g.} number of requested call establishments in IMS).

\subsection{Elastisity-QoS Trade-off}
\label{sec:tradeoff}

The advantages of resource sharing has drawn attention to design of elastic architectures to carry telecommunication applications.
In such architectures, low amount of available computing resources may result in bad or unacceptable QoS level.
Especially, when the time required for creating (or allocating) new computing nodes is long or the increasing rate of the load on the system is high, the performance of telecommunication applications may be affected.

To address this issue, a safe bound is usually considered between the number of allocated pouches and the number of required pouches to handle the current load.
This safe bound will be able to accommodate the new incoming requests (\textit{i.e.} new call establishment requests from UEs) until new pouches are allocated for the IMS application layer.

The size of this safe bound (more precisely the number of extra-allocated pouches) specifies the chance of being overloaded (and hence receiving a busy signal for a new call).
Specifically, a small number of extra-allocated pouches will increase the likelihood of falling in a busy situation.
On the opposite side, picking a large number of extra-allocated pouches is inefficient and move us far from having an ideal elastic deployment.
In summary, there is a trade-off between the experienced QoS and the level of elastisity that we can achieve and it is controlled by allocating an appropriate number of extra-allocated pouches.
In practice, this task can be done by studying the statistics of the subscriber requests and analyzing the latency of different parts of the implementation.

\section*{Acknowledgments}
This work was supported by Ericsson Canada Inc and MITACS Canada through a MITACS Accelerate partnership program between Ericsson Canada Inc and McGill University.
We would also like to thank the team of dedicated researchers, Marc-Olivier Arsenault, Gordon Bailey, Mario Bonja, Leo Collard, Philippe Habib, Olivier Lemelin, Maxime Nadeau, Fakher Oueslati and Joseph Siouffi, for their endeavoring effort in developing our proof of concept implementation.

\bibliographystyle{ieeetr}
\bibliography{ref}

\end{document}